# Evidence for interfacial superconductivity in a bi-collinear anti-ferromagnetically ordered FeTe monolayer on a topological insulator

S. Manna[1], A. Kamlapure[1], L. Cornils[1], T. Hänke[1], E. M. J. Hedegaard[2], M. Bremholm[2], B. B. Iversen[2], Ph. Hofmann[3], J. Wiebe[1,*] and R. Wiesendanger[1]

[1]Department of Physics, University of Hamburg, Jungiusstrasse 11, D-20355 Hamburg, Germany

[2]Center for Materials Crystallography, Department of Chemistry and iNANO, Aarhus University, Langelandsgade 140, DK-8000 Aarhus C, Denmark

[3]Department of Physics and Astronomy, Interdisciplinary Nanoscience Center, Aarhus University, DK-8000 Aarhus C, Denmark

[*]Corresponding author: jwiebe@physnet.uni-hamburg.de

**The discovery of high-temperature superconductivity in Fe-based compounds [1,2] has triggered numerous investigations on the interplay between superconductivity and magnetism [3] and, more recently, on the enhancement of transition temperatures through interface effects [4]. It is widely believed that the emergence of optimal superconductivity is intimately linked to the suppression of long-range antiferromagnetic (AFM) order, although the exact microscopic picture of this relationship remains elusive [1] due to the lack of data with atomic spatial resolution [5-7]. Here, we present a spin-polarized scanning tunneling spectroscopy (SP-STS) study of ultrathin $FeTe_{1-x}Se_x$ (x = 0, 0.5) films grown on prototypical Bi-based bulk topological insulators. Surprisingly, we find an energy gap at the Fermi level indicating superconducting correlations up to $T_c$ ~ 6 K for one unit cell thin FeTe layers grown on $Bi_2Te_3$ substrates, in contrast to the non-superconducting FeTe bulk compound [8]. Moreover, SP-STS reveals that the energy gap spatially coexists with bi-collinear AFM order. This finding opens novel perspectives for theoretical studies of competing orders in Fe-based superconductors as well as for experimental investigations of exotic phases in heterostructures of topological insulators and superconducting layers.**

Topological insulators (TI) and interfacial superconductors (SC) are both topics of intense current interest in modern condensed matter physics [4,9]. The combination of both materials is expected to reveal novel physics such as, e.g., Majorana Fermions arising in heterostructures of TIs and s-wave SCs by including magnetic fields [10]. Most experimental studies of such heterostructures have concentrated on TI films grown on superconducting bulk substrates [11]. Alternatively, SC/TI heterostructures could be realized by growing superconducting films on top of pristine bulk TIs.

The recent finding that the superconducting transition temperature $T_c$ can be pushed above 100 K in an FeSe unit cell (UC) thin film grown epitaxially on $SrTiO_3$ [12-14] spurred numerous investigations aiming at an exact understanding of how the superconductivity in transition metal mono- and di-chalcogenides evolves from bulk to ultra-thin films. Bulk FeSe crystals exhibit a $T_c$ of no higher than 8 K [15]. The significant interfacial enhancement of $T_c$ in UC FeSe on $SrTiO_3$ as well as in similar systems, such as FeSe on $BaTiO_3$ [16] and $FeTe_{1-x}Se_x$ on $SrTiO_3$ [17], lead to a revival of the idea that tailoring the electron pairing by interfacing a SC with another material can be used to achieve high-temperature SC [4]. Interestingly, SC can even emerge by interfacing two insulators [7] or a topological insulator ($Bi_2Te_3$) with a non-SC Fe chalcogenide (FeTe) [18,19]. It is therefore of great importance to understand how different substrates affect the electronic, spin-dependent and superconducting properties of single layer Fe-based superconductors.

The latter studies [18,19] raised the fundamental question of what happens to the bi-collinear AFM structure of FeTe [20] once an interfacial SC state is established in contact with $Bi_2Te_3$. In the work of He *et al.* [18], it was proposed that the topological surface state may dope the FeTe and suppress the long-range AFM order at the interface, thereby inducing the observed interfacial SC. By a low-temperature spin-polarized scanning tunneling spectroscopy (SP-STS) investigation of the correlation between the film structure, the local electronic, superconducting and spin-dependent properties, we show here that, in contrast to this assumption, the energy gap at the Fermi-level in UC thin FeTe films epitaxially grown on $Bi_2Te_3$(111) can spatially coexist on the atomic scale with long-range bi-collinear AFM order.

We have prepared ultrathin epitaxial FeTe films on $Bi_2Te_3$ single-crystals by *in-situ* deposition of Fe on the TI substrates under ultra-high vacuum (UHV) conditions, followed by an *in-situ* annealing treatment (see Methods section). Figure 1(A) depicts a constant-current STM topographic image of such a film with FeTe islands of different thicknesses, together with regions of the bare substrate. We distinguish between these different areas based on the measured step heights and differences in the observed atomic-scale surface structures. The line profile displayed in Fig. 1(B) taken across three different terraces separated by two steps reveals a step height $h$ ("ab") of $h \sim 0.65\pm0.05$ nm, roughly equal to the *c*-axis lattice constant (0.62 nm) of bulk FeTe, while the right step "bc" is considerably smaller $h \sim 0.35\pm0.05$ nm. The smaller step results from an FeTe UC being embedded in the topmost quintuple layer of the substrate, most probably as sketched in Fig. 1(B). Figure 1(E) presents an atomically resolved STM topographic image of the substrate area ($Bi_2Te_3$), which reveals the hexagonal close-packed atomic lattice (see inset FFT image) of top Te atoms with a spacing of ~ 0.44 nm. A topographic STM image of the top UC layer of FeTe, with an atomic contrast determined by the topmost chalcogenide (Te) layer (Fig. 1(F)), reveals that the *a*-axis of the FeTe is aligned parallel to one of the closed packed directions of the substrate Te atoms. The period of the atomic lattice along the *a*-axis direction is ~ 0.38 nm, which is consistent with the Te-Te atomic distance on the surface of bulk FeTe [21]. The period along the *b*-axis is slightly smaller ($a/b \sim 1.04$), which might be partly attributed to the transition into the orthorhombic/monoclinic phase, known for bulk FeTe [8], but could be enhanced by uniaxial strain as a result of the heteroepitaxial growth. This holds for the embedded UC FeTe as well as for thicker layers (Supplementary Fig. S1). However, there is no overall periodic height modulation of the UC FeTe layers, and only in some areas of the thinnest islands irregular wrinkles are observed. This indicates a rather weak interaction of the FeTe layer with the substrate, in contrast to what has been observed for FeSe grown on $Bi_2Se_3$ [22]. The Fourier transform of the topographic image (inset of Fig. 1(F)) clearly shows the peaks associated with the almost square Te atomic lattice, with the spots along the *a*-direction ($\mathbf{q}_{Te}^a$) slightly more intense than those along the *b*-direction ($\mathbf{q}_{Te}^b$) [20]. Despite the different lattice symmetries (six-fold for $Bi_2Te_3$ and four-fold for FeTe) with respect to the *c*-axes directions and the relatively large lattice mismatch, the heteroepitaxial growth is of very high quality, resulting in an atomically sharp and defect-free interface. In contrast to the surface of bulk FeTe [20], the surfaces of our ultrathin FeTe films do not show any excess Fe atoms, indicating a stoichiometric FeTe layer.

High-quality, one UC thin $FeTe_{0.5}Se_{0.5}$ films (Supplementary Fig. S2) with well-defined stoichiometry were epitaxially grown on $Bi_2Te_{1.8}Se_{1.2}(111)$ substrates using a similar preparation method as for the FeTe films. Atomically resolved topographic STM images of the cleaved $Bi_2Te_{1.8}Se_{1.2}$ substrate surface (Fig. 1(G)) show Se- and Te-atoms in the topmost layer of the surface appearing with slightly darker and brighter contrast, respectively, with a frequency in accordance with the stoichiometry [23]. A typical topographic STM image of a one UC thin $FeTe_{0.5}Se_{0.5}$ film grown on top (Fig. 1(H)) reveals an ordered square lattice of Se/Te atoms with about half of the atoms appearing brighter (52%, Te-sites) while the other half appears darker (48%, Se-sites), with a height difference of 65±15 pm, close to the value found for similarly doped FeTe$_{1-x}$Se$_x$ samples [17,24].

To characterize the electronic structure of these films, STS measurements of the differential tunneling conductance (d$I$/d$U$) as a function of the applied bias voltage were performed, reflecting the sample's local density of states (LDOS). Figure 2(A) shows single tunneling spectra measured on the one UC thin FeTe$_{0.5}$Se$_{0.5}$ film and on the Bi$_2$Te$_{1.8}$Se$_{1.2}$ substrate at a temperature of $T$ = 1.1 K. The FeTe$_{0.5}$Se$_{0.5}$ film exhibits an overall U-shaped d$I$/d$U$ spectrum with an almost vanishing conductance value at the Fermi level ($E_F$) and an energy gap of $\Delta$ = 2.5 meV, defined by half the distance between the two sharp coherence peaks in Fig. 2(A). This measured gap value is of the same order of magnitude as measured on the surface of the corresponding superconducting bulk material [25]. In order to determine the critical temperature $T_c$ of our FeTe$_{0.5}$Se$_{0.5}$ films, we performed temperature-dependent measurements of d$I$/d$U$ from 1.1 K to 11 K (see Fig. 2B). As expected, the observed energy gap in d$I$/d$U$ becomes shallower with increasing temperature and eventually disappears at about 11 K. The energy gap values for different temperatures (shown in Fig. 2C) were extracted from fits of the background-corrected and symmetrized d$I$/d$U$ spectra (for data processing see Supplementary Fig. S3) employing the Dynes density of states, $N(E) = N_n(E_F) \cdot \Re\left[\frac{|E|-i\Gamma}{\sqrt{(|E|-i\Gamma)^2-\Delta^2(T)}}\right]$, with the density of states in the normal state at the Fermi energy $N_n(E_F)$ and the real part $\Re$, where the parameter $\Gamma$ accounts for the lifetime broadening of the quasiparticles [26]. The fitting of the temperature dependence of the gap in the framework of the Bardeen–Cooper–Schrieffer (BCS) theory [27] yields an energy gap at 0 K of $\Delta_0$ = 2.0 meV and an extracted $T_c$ = 11 K, similar to the values that have been extracted from STS on the corresponding superconducting bulk materials [25,28]. The temperature-dependence of the d$I$/d$U$ spectra as well as the derived $T_c$ value provide strong evidence that the observed gap feature arises from superconductivity in the ultrathin $FeTe_{0.5}Se_{0.5}$ film grown on $Bi_2Te_{1.8}Se_{1.2}$.

Now we turn to the investigation of the electronic and magnetic structure of ultrathin FeTe films grown on $Bi_2Te_3$. Figure 2(D) shows a series of spatially averaged d$I$/d$U$ spectra measured at $T = 1.1$ K on the $Bi_2Te_3$ substrate as well as on ultrathin FeTe films of different thicknesses. For both embedded UC and top UC FeTe layers a reduced d$I$/d$U$ at $E_F$ with diffuse peaks at energies of ±1 meV to ±2 meV are found with a gap width and occurrence of the peaks varying with position (Supplementary Fig. S4). A gap of this size and temperature dependence (shown below) has not been observed at the surface of bulk FeTe [29,30] or for FeTe single layers grown on $SrTiO_3$ [17]. The Fermi level gap gradually vanishes for increased thicknesses of FeTe layers on $Bi_2Te_3$, e.g. two UC thin FeTe films show a gap with a significantly reduced depth at $E_F$. This behavior suggests that the $FeTe/Bi_2Te_3$ interface plays a significant role for the observed gap structure. In order to reveal the origin of the energy gap in one UC thin FeTe films grown on $Bi_2Te_3$, we thoroughly investigated the temperature- and field-dependence of the d$I$/d$U$ spectra. Figure 2(E) shows a series of background-corrected and symmetrized temperature-dependent d$I$/d$U$ spectra for top UC FeTe on $Bi_2Te_3$. The tunneling spectrum at $T = 1.1$ K exhibits a pair of diffuse peaks at $U \sim 2$ meV together with a gap region of reduced LDOS at $E_F$. We observe that the gap depth and peak height in d$I$/d$U$ decrease with increasing temperature and finally vanish above $T = 6$ K, very reminiscent of a superconducting transition. In order to test this hypothesis, the temperature dependent spectra were fitted to a Dynes density of states (Fig. 2(E)). Here we note that a considerable lifetime broadening $\Gamma$ had to be assumed (Supplementary Fig. S5). The temperature dependence of the fitted energy gap values (Fig. 2(F)) resembles that of a superconducting phase transition (cf. Fig. 2(C)). Kondo correlations, which were shown to cause dip features at $E_F$ for heavy Fermion systems [31] and might be present on the Fe lattice as well, are ruled out, as they would show a different temperature dependency of the gap depth (Supplementary Fig. S6). The spectral features therefore strongly suggest superconducting correlations of the UC FeTe layer. In order to extract the corresponding $T_c$, we fit the temperature dependence of the gap in the framework of the BCS theory (Fig. 2(F)), resulting in $T_c \sim 6.5$ K. Together with the zero temperature gap value of $\Delta_0 = 1.05$ meV extrapolated from the BCS fit we deduce a ratio of $2\Delta_0/k_B T_c \sim 3.8$ for one UC thin FeTe on $Bi_2Te_3$ which is comparable to the value of 4.2 we determine for the $FeTe_{0.5}Se_{0.5}$ thin film on $Bi_2Te_{1.8}Se_{1.2}$, but somewhat larger than typical values of 2.5 to 3.2 for bulk $FeTe_{1-x}Se_x$ samples [25,28]. We note that the embedded UC of FeTe also shows a similar temperature dependent behavior (Supplementary Fig. S7). Our results are consistent with the transport measurements of $Fe_{1+x}Te$ films grown on $Bi_2Te_3$ [19] where evidence for superconductivity with a $T_c \sim 12$ K was reported. We do not observe any significant effect on the energy gap in a magnetic field up to $B = 2.8$ T applied perpendicular to the sample plane (Supplementary Fig. S8). This suggests that the robustness of the pairing against an external magnetic field is anomalously high, similar to what has been observed in magnetotransport measurements [18] and other Fe based superconducting thin films [12].

Having established clear evidence for superconducting correlations in single UC FeTe layers grown on $Bi_2Te_3$ below 6 K, we now focus in more detail on the spatial behavior of these correlations at the lateral interface of the FeTe layer and the $Bi_2Te_3$ substrate. In particular, we measured the spatial variation of tunneling spectra across a step edge from the embedded UC FeTe layer to the $Bi_2Te_3$ substrate (Fig. 3(A,B)). The resulting tunneling spectra across the interface along the line shown in Fig. 3(A) obtained at $T = 1.1$ K (Fig. 3(C)) again show the presence of the gap due to superconducting correlations on top of the FeTe layer, which vanishes on the topological insulator substrate. To quantify the decay length of the superconducting correlations in our heterostructure system, we have fitted the spatial dependence of the gap area with an exponential decay function (Fig. 3(D)), resulting in a decay length of about $\xi = 8.9$ Å at $T = 1.1$ K. This decay length is a measure for the coherence length of the Cooper pairs, which is obviously similarly small as, e.g., in bulk Fe Te$_{0.6}$Se$_{0.4}$ [28].

Our results on epitaxially grown FeTe layers on Bi$_2$Te$_3$ confirm that the previously reported 2D superconductivity in FeTe/Bi$_2$Te$_3$ heterostructures [18,19] is associated with the presence of a superconducting layer of FeTe located at the interface. We now return to the central question of whether the usual bi-collinear AFM order of bulk FeTe [20] is suppressed in the FeTe layer interfacing Bi$_2$Te$_3$, as proposed by [18], or whether it can co-exist with the superconducting correlations we observed.

In order to check this assumption we simultaneously characterized the local superconducting correlations and the atomic-scale spin structure of the ultrathin FeTe films grown on Bi$_2$Te$_3$ by SP-STS [32]. By using a spin-polarized bulk Cr tip [33], we excluded the disturbing influence of a local magnetic stray field on the sample while being sensitive to its out-of-plane surface spin component. Figure 4(A) represents the spin-resolved constant-current image of a single UC thin embedded FeTe layer on Bi$_2$Te$_3$ as obtained in an external magnetic field of $B = +1$ T applied perpendicular to the sample surface. The SP-STM image shows a clear stripe-like pattern, superimposed on the square atomic lattice of surface Te atoms, which is significantly different from the STM topography obtained with a non-magnetic PtIr tip as presented in Fig. 1(F). The Fourier transform (inset of Fig. 4(A)) of the SP-STM image reveals an additional pair of peaks along the *a*-direction with a wave vector $\mathbf{q}_{AFM} = 1/2\ \mathbf{q}_{Te}^a$ that is half as long as that of the atomic lattice Bragg peaks. This unidirectional modulation with a characteristic periodicity of λ = 2a = 7.5 Å (see Fig. 4(E)) has been attributed to a direct observation of bi-collinear AFM order of the Fe atoms below the surface of bulk $Fe_{1+y}Te$ using SP-STM [20]. To confirm the magnetic origin of the observed superstructure with twice the atomic lattice periodicity, we performed an SP-STM measurement at the same location with an oppositely aligned effective spin at the Cr tip's front end by reversing the magnetic field to B = -1 T (see Fig. 4(B)). Now, the 2a superstructure reveals a phase shift (indicated by an arrow) of one atomic lattice unit as expected for a magnetic origin of the contrast. The maximum spin contrast appears between every second Fe lattice site along a diagonal of the surface unit cell being located between two neighboring Te sites. This can be explained by the fact that spin-polarized tunneling primarily results from the 3d states of Fe being located below the top Te layer. It is also clearly visible from the line profiles of Fig. 4(E) along the *a*-direction, which were extracted from Figs. 4(A) and 4(B), and by the 2a periodicity along the *a*-direction in the out-of-plane spin-polarization distribution of the FeTe layer (Fig. 4(C)) obtained by subtracting the SP-STM images of Figs. 4(A) and 4(B). These observations prove that single UC FeTe films on $Bi_2Te_3$ exhibit a bi-collinear AFM structure, similar to the one known for bulk FeTe [20], but with a *non-vanishing* out-of-plane surface spin component. Note, that the same bi-collinear AFM structure was also observed for the thicker islands in our samples (Supplementary Fig. S9). To our knowledge this is the first direct real-space observation of magnetic order in an Fe-chalcogenide compound at the single unit-cell thickness limit.

Figure 4(F) shows the intensity ratios of the AFM peaks (at $q_{AFM}$) to the Bragg peaks (at $q_{Te}^a$) of the FFTs of SP-STM data taken from top UC FeTe as a function of temperature. It clearly demonstrates that the bi-collinear AFM order is largely unaffected by the disappearance of the superconducting correlations occurring at 6.5 K (Fig. 2(F)). Therefore our experimental results do not provide any evidence that one kind of order emerges at the expense of the other, nor does the data provide any indication for a microscopic phase separation into regions with superconducting and magnetic order. Our findings for single UC thin FeTe layers on $Bi_2Te_3$ therefore challenge the common belief that optimal superconducting pairing sets in when long-range AFM order is suppressed in the parent compound.

In summary, we have explored the electronic and magnetic structure in a new class of systems, i.e. heterostructures consisting of ultrathin Fe-chalcogenide layers of the type $FeTe_{1-x}Se_x$ (x=0, 0.5) on Bi-based TI substrates. For $FeTe/Bi_2Te_3$ heterostructures, our atomic-scale spin-resolved tunneling spectroscopy measurements provide evidence for the coexistence of superconducting correlations with a $T_c \sim 6.5$ K and bi-collinear AFM order in the one UC FeTe films. We finally compare the wavelength of the AFM order of $\lambda = 2a = 7.5$ Å with the size of the Cooper pairs, inferred from the coherence length, of $\xi = 8.9$ Å, giving $\xi/\lambda \sim 1.2$. The electron distance in the pairs is rather small, which is typical for Fe-based SCs, and apparently just large enough such that the effective Zeeman field induced by the AFM order cancels out along the length of the Cooper pair. The relative sizes of $\lambda$ and $\xi$ therefore might be crucial for the coexistence of pairing and long range AFM order in this material. Our surprising findings may stimulate further theoretical studies on the relationship between superconductivity and AFM order in Fe based SCs.

# Methods

## Samples

Bulk topological insulator (TI) single crystals of $Bi_2Te_3$ and $Bi_2Te_{1.8}Se_{1.2}$ used in this study as substrates were synthesized using a Stockbarger method and have been well characterized using angle-resolved photoemission spectroscopy, powder X-ray diffraction, inductively coupled plasma atomic-emission spectroscopy, and potential Seebeck microprobe measurements [23,34]. Fe-chalcogenide film preparation and characterization were carried out in a UHV system with a base pressure below $1 \times 10^{-10}$ mbar. The TI crystals were cleaved *in-situ* under UHV conditions and high quality $FeTe_{1-x}Se_x$ (x = 0, 0.5) thin films were prepared by depositing 0.5–1 ML Fe at 300 K on-top of $Bi_2Te_3$ and $Bi_2Te_{1.8}Se_{1.2}$ substrates, respectively, using molecular beam epitaxy, followed by annealing up to a maximum temperature of 315 °C for 15 minutes. Fe deposited on $Bi_2Te_3$ reacts with the substrate upon annealing, most likely via a substitutional process of Bi by Fe [35], leading to a high-quality defect-free FeTe film.

## Experimental techniques

All scanning tunneling microscopy and spectroscopy (STM/STS) experiments were performed with an STM in UHV at temperatures between 1.1 K and 14 K [26]. A magnetic field $B$ of up to 3 T can be applied perpendicular to the sample surface. Topography images were obtained in constant-current mode with stabilization current $I_s$ and bias voltage $U$ applied to the sample. STS data were obtained using a lock-in technique to record the differential tunneling conductance $dI/dU$ by adding an AC modulation voltage $U_{mod}$ (given in rms) to the bias voltage, after stabilizing the tip at $I_s$ and $U$, switching off the feedback, and ramping the applied bias $U$. We used cut PtIr or electrochemically etched W tips (both *in situ* flashed) for spin-averaged imaging and spectroscopy measurements. For spin-resolved measurements, we used bulk Cr tips, which were prepared by electrochemical etching followed by a high voltage field emission treatment using W(110) or Ta(001) as a substrate.

## Acknowledgments:

We are indebted to Alexander Balatsky, Christopher Triola, Tim O. Wehling and Udai R. Singh for valuable discussions. This work has primarily been supported by the ERC Advanced Grant ASTONISH (No. 338802). L.C., Ph.H. and J.W. acknowledge partial support through the DFG priority program SPP1666 (grant No. WI 3097/2), and T.H. acknowledges support by the DFG under grant No. HA 6037/2-1. We thank the Aarhus University Research Foundation for supporting the bulk crystal growth.

## Author contributions:

S.M., A.K. and L.C. contributed equally to this work. S.M., J.W., T.H. and R.W. designed the experiment. S.M., A.K. and L.C. have grown and characterized the thin film samples and performed the STM/STS and SP-STS experiments. S.M., A.K. and L.C. analyzed the data. S.M, R.W. and J.W. wrote the manuscript. E.M.J.H., M.B., B.B.I. and Ph.H. have grown and characterized the TI single crystals. All authors discussed the results and contributed to the manuscript.

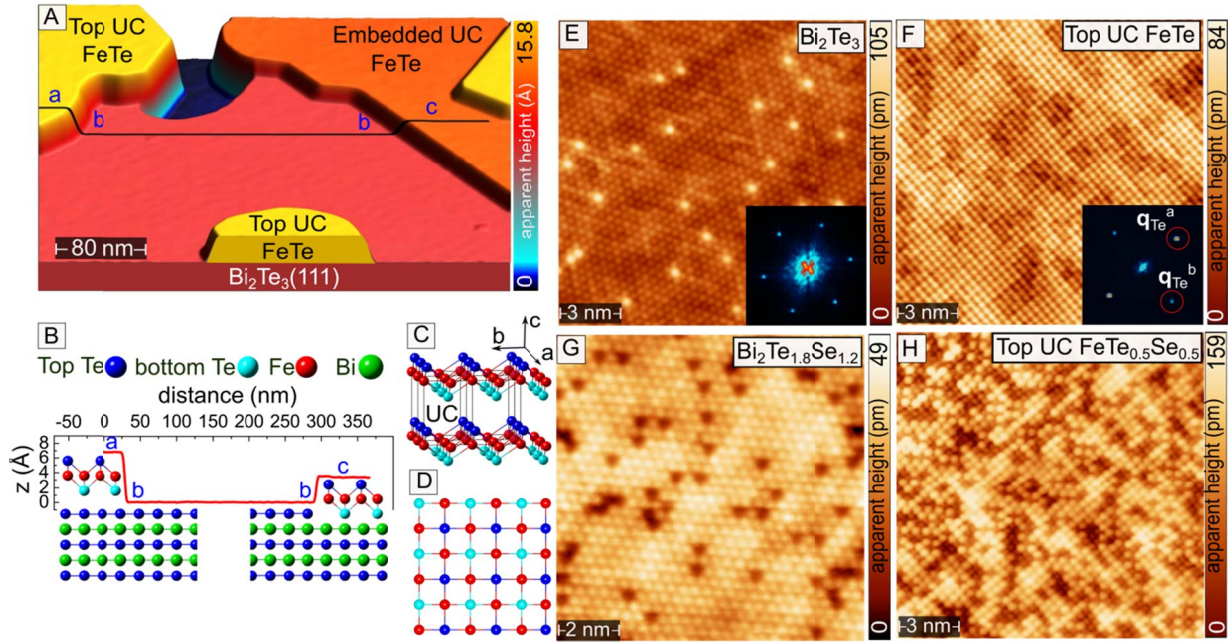

**Fig. 1 STM topography of FeTe$_{1-x}$Se$_x$ (x = 0, 0.5) ultrathin films grown on Bi-based TI substrates**. **(A)** Large-scale (382 nm x 375 nm) constant-current image ($U$ = 500 mV, $I_s$ = 20 pA) of one unit-cell (UC) FeTe grown on a Bi$_2$Te$_3$ substrate. **(B)** Measured height profile along the black line (a-c) in **(A)** revealing embedded ($h$ = 0.35 nm) and top UC ($h$ = 0.68 nm) FeTe, consistent with the suggested cross-sectional schematic illustration of the FeTe/Bi$_2$Te$_3$ heterostructure (not to scale). The schematic crystal structure of an FeTe UC is additionally shown in side- **(C)** and top-view **(D)**. **(E)** Atomically resolved STM image (15 nm x 15 nm) of a Bi$_2$Te$_3$ substrate region ($U$ = -200 mV, $I_s$ = 100 pA), and **(F)** of a top UC thin FeTe layer ($U$ = 140 mV, $I_s$ = 100 pA). Fast Fourier transforms (FFT, image sizes 0.75 Å$^{-1}$) of the images are displayed in the insets of (E) and (F), revealing the Bragg peaks associated with the top Te atomic lattice at $\mathbf{q}_{Te}^a$ and $\mathbf{q}_{Te}^b$. **(G)** Atomically resolved STM image of the Bi$_2$Te$_{1.8}$Se$_{1.2}$ substrate ($U$ = 50 mV, $I_s$ = 1 nA) and **(H)** of one UC FeTe$_{0.5}$Se$_{0.5}$ layer grown on top ($U$ = 100 mV, $I_s$ = 740 pA).

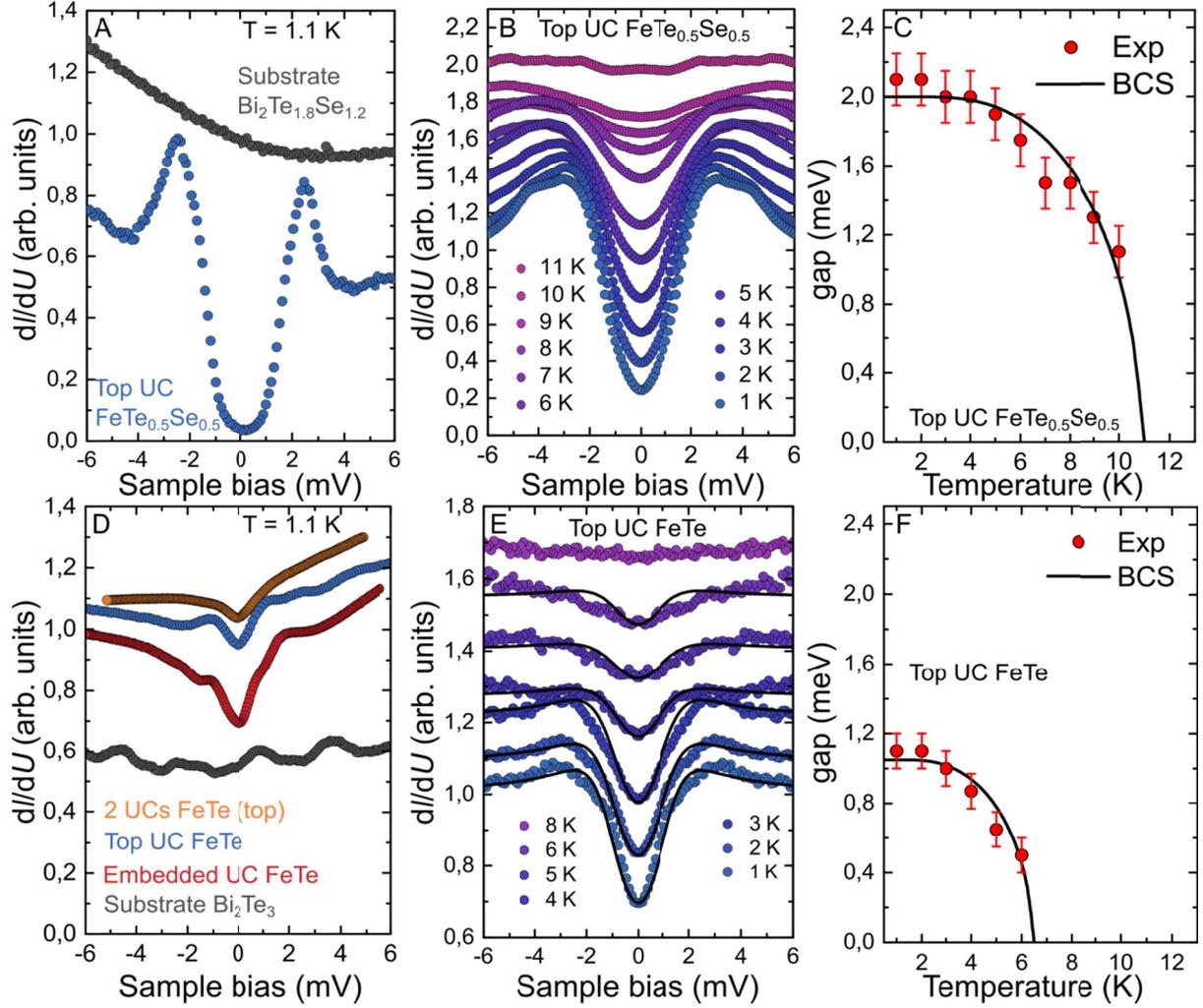

**Fig.2 Thickness and temperature dependence of the tunneling conductance spectra on FeTe$_{1-x}$Se$_x$ (x=0, 0.5). (A)** d$I$/d$U$ spectra measured on one UC thin FeTe$_{0.5}$Se$_{0.5}$ and on the Bi$_2$Te$_{1.8}$Se$_{1.2}$ substrate at $T$ = 1.1 K ($U$ = 10 mV, $I_s$ = 400 pA, $U_{mod}$ = 0.1 mV). **(B)** Evolution of the d$I$/d$U$ spectra on one UC thin FeTe$_{0.5}$Se$_{0.5}$ as a function of temperature ($U$ = 10 mV, $I_s$ = 200 pA, $U_{mod}$ = 0.15 mV, spectra vertically offset for clarity). The spectra are spatially averaged along a line of length 1 nm, background-corrected by division through the data measured above $T_c$ (at $T$ = 14 K), and finally symmetrized with respect to $U$ = 0 V (Supplementary Fig. S3). **(C)** The energy gap of one UC thin FeTe$_{0.5}$Se$_{0.5}$ as a function of temperature. Markers show the gaps as extracted from Dynes fits (not shown) to the data in (B), and the solid line shows the fitting to the BCS gap function. **(D)** Typical spatially averaged d$I$/d$U$ spectra measured on ultrathin FeTe films of different thicknesses and on the Bi$_2$Te$_3$ substrate at $T$ = 1.1 K. The spectra have been normalized to the d$I$/d$U$ value at +5 mV and are vertically offset (substrate: offset -0.4, $U$ = 10 mV, $I_s$ =170 pA, $U_{mod}$ = 0.2 mV; embedded UC: offset +0.1, $U$ = 6 mV, $I_s$ =300 pA, $U_{mod}$ = 0.2 mV; top UC: offset +0.2, $U$ = 10 mV, $I_s$ =600 pA, $U_{mod}$ = 0.25 mV; 2 UCs: offset +0.3, $U$ = 5 mV, $I_s$ =300 pA, $U_{mod}$ = 0.2 mV). **(E)** Spatially (0.5 nm x 0.5 nm) averaged d$I$/d$U$ spectra measured on top of a one UC thin FeTe layer at temperatures from 1.1 K to 8 K ($U$ = 10 mV, $I_s$ = 200 pA, $U_{mod}$ = 0.15 mV, spectra vertically offset for clarity). The spectra are background-corrected by division through the data measured at $T$ = 8 K, and finally symmetrized with respect to $U$ = 0 V. The lines are fits to Dynes functions with an energy broadening as given in Supplementary Fig. S6. **(F)** Energy gaps Δ, as obtained from the fits in (E), as a function of temperature. The solid line represents the temperature variation of Δ as fitted by BCS theory.

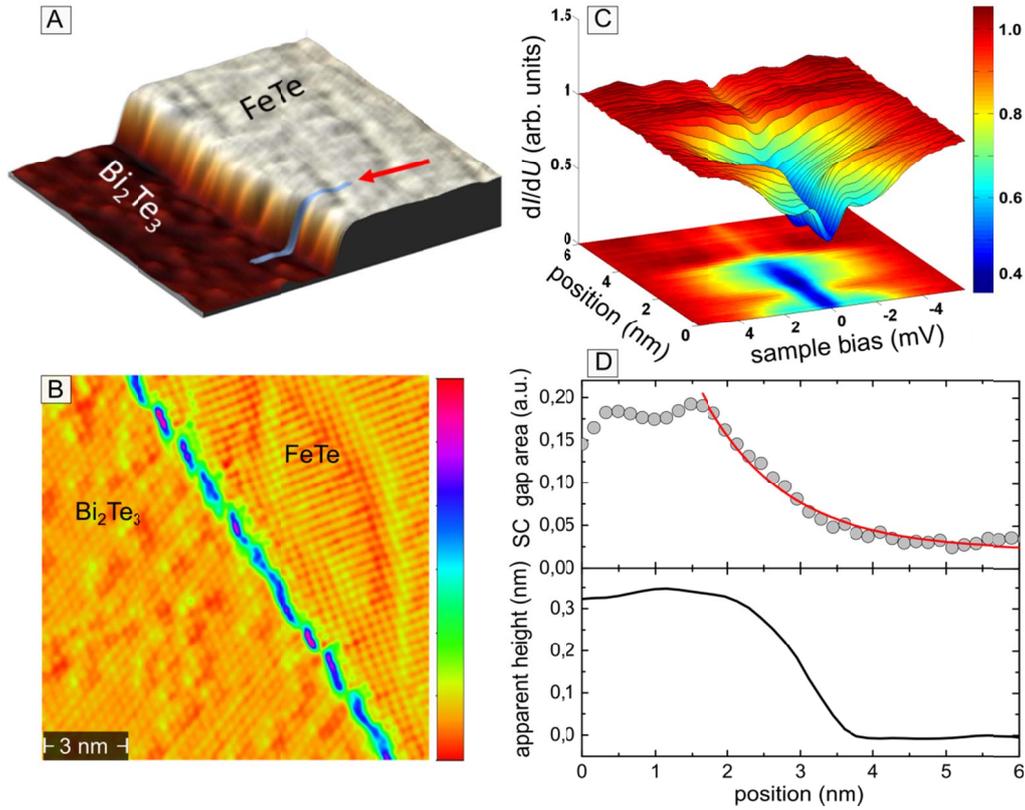

**Fig.3 Spatial distribution of energy gap due to SC correlations across a lateral interface.** **(A)** STM topography (25 nm x 25 nm) showing a Bi$_2$Te$_3$ terrace (left) and an embedded UC thin FeTe layer (right) grown on top ($U$ = 100 mV, $I_s$ = 50 pA) **(B)** Atomically resolved STM topography (15 nm x 15 nm, differentiated with respect to horizontal axis) of a similar area ($U$ = 50 mV, $I_s$ = 50 pA). **(C)** 2D and 3D representation of normalized and symmetrized d$I$/d$U$ spectra taken across the step along the line indicated in **(A)** in the direction of the arrow. **(D)** Evolution of the gap (top, markers), determined from the spectra in (B) by measuring the area enclosed by the gap within a ±1 meV voltage window, and topographic height (bottom) taken along the same line in (C). The solid line in the gap area plot shows a fit to an exponential decay resulting in a decay length of ξ = 8.9 Å.

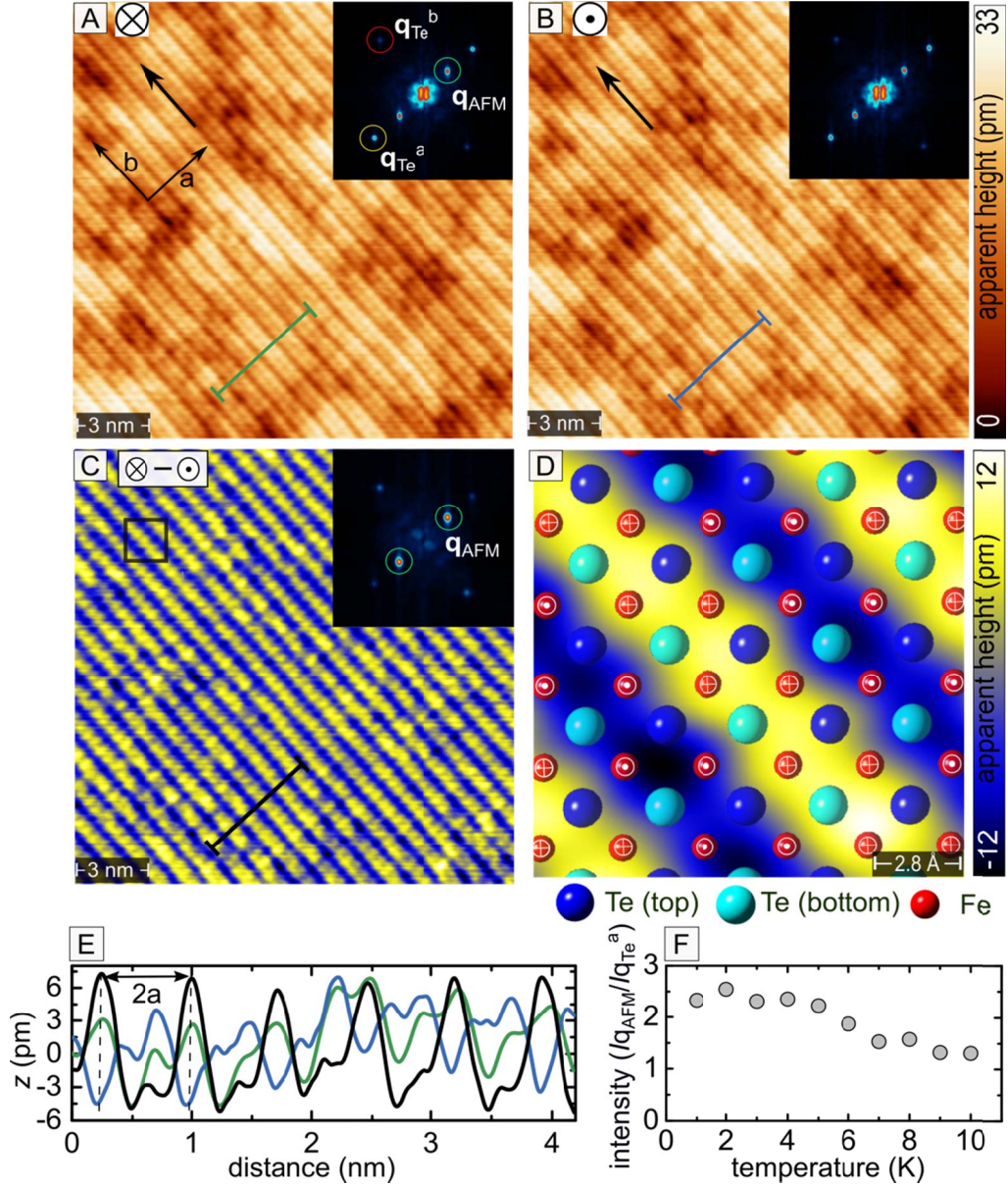

**Fig.4 Bi-collinear AFM order in embedded and top UC FeTe revealed by SP-STM. (A)** SP-STM image recorded at $B = +1$ T showing spin order of one UC FeTe at $T = 1.1$ K using an out-of-plane spin-sensitive bulk Cr tip. **(B)** SP-STM image of the same area acquired at $B = -1$ T with an oppositely spin-oriented tip ($U = 300$ mV, $I_s = 220$ pA). The insets in (A) and (B) show FFTs of the corresponding SP-STM images, where the peaks marked by red and yellow circles are associated with the top Te atomic square lattice (Bragg peaks) at $q_{Te}^a$ and $q_{Te}^b$ (image sizes of all FFTs is 0.55 Å$^{-1}$)). Besides the Te Bragg peaks, the FFTs reveal pairs of additional strong peaks associated with AFM order at $q_{AFM} = 1/2\ q_{Te}^a$. **(C)** Spin asymmetry map obtained by subtraction of images (A) and (B), revealing bi-collinear AFM order in the one UC thin FeTe layer. Inset: FFT pattern of the difference image highlighting the peaks in reciprocal space resulting from AFM spin order. **(D)** Model representation of the atomic structure and of the out-of plane components of magnetic moments superimposed on the zoomed-in spin asymmetry image of (C). **(E)** Measured profiles along identical lines drawn in images (A)-(C), showing the periodicity of AFM order in real space, i.e. $\lambda = 2a$. **(F)** Temperature dependent spin contrast of top UC FeTe quantified by the FFT intensity ratios of the $\mathbf{q}_{AFM}$ and $\mathbf{q}_{Te}^{a}$ peaks (see Supplementary Fig. S10 for the corresponding SP-STM images).

# Supplementary information

**Supplementary Figures:**

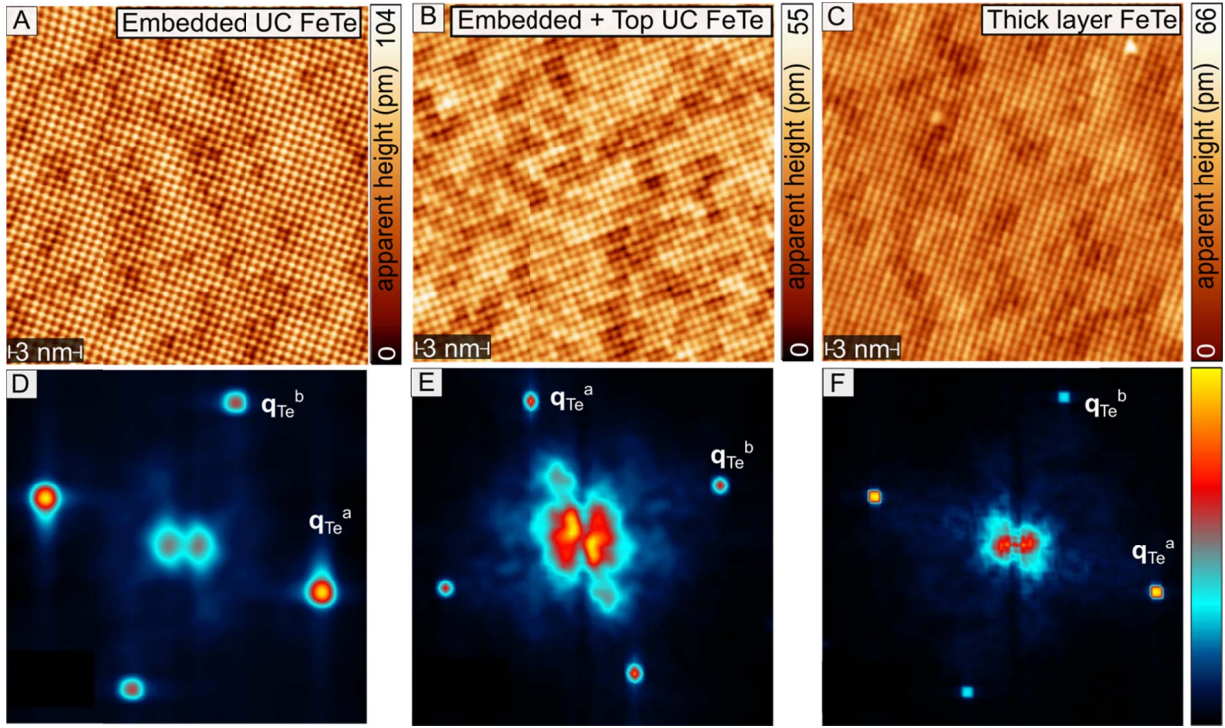

**Fig. S1: Atomically resolved STM topographs (A-C) of FeTe films of different thicknesses, as indicated, grown on Bi$_2$Te$_3$, and corresponding FFT images (D-F).** The ratios of the lattice constants in *a*- and *b*-direction, *a/b*, derived from the atomically resolved topographs and their FFTs are: **(A,D)** *a/b* ~ 1.05, **(B,E)** *a/b* ~ 1.01, **(C,F)** *a/b* ~ 1.07. Measurement parameters of the topographs: (**A**) $U$ = 100 mV, $I_s$ = 100 pA, (**B**) $U$ = -15 mV, $I_s$ = 300 pA, (**C**) $U$ = 280 mV, $I_s$ = 100 pA. Image size of all FFTs is 0.65 Å$^{-1}$.

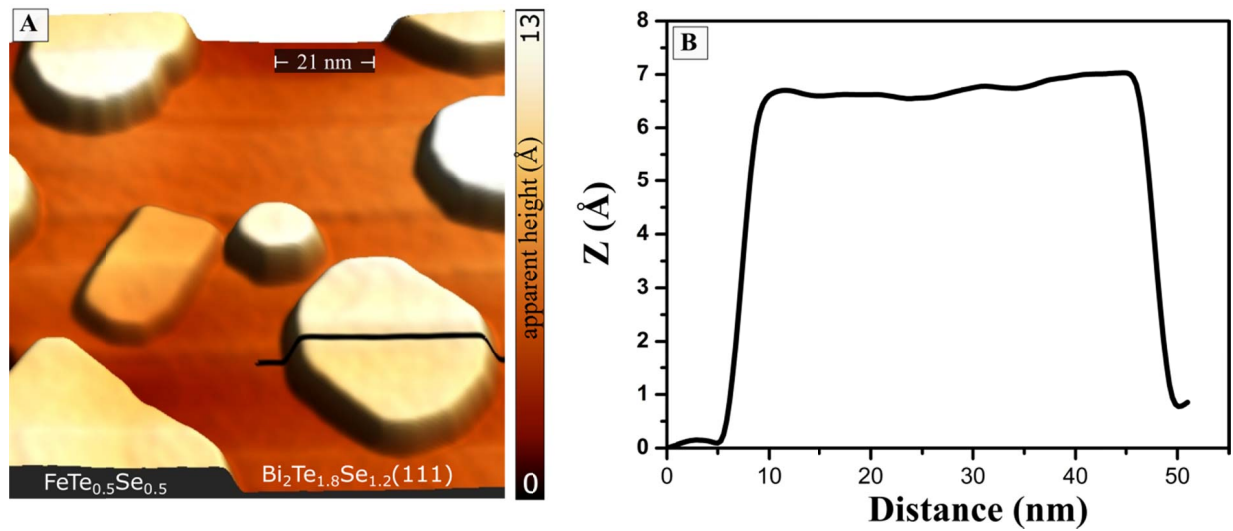

**Fig. S2: Overview topograph of a thin film of FeTe$_{0.5}$Se$_{0.5}$ grown on Bi$_2$Te$_{1.8}$Se$_{1.2}$.** (**A**) The STM topograph shows one embedded UC thick island as well as several top UC thick islands. The height profile of one of the top UC islands along the indicated line is shown in (**B**). Measurement parameters: $U = 100$ mV, $I_s = 150$ pA.

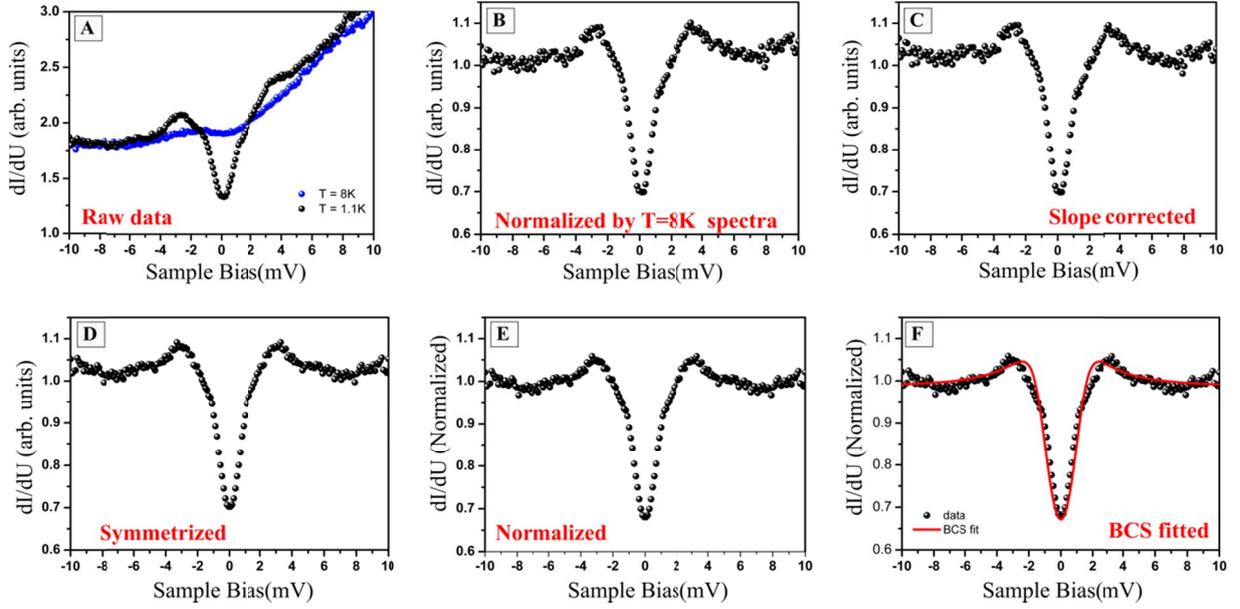

**Fig. S3: Description of the data processing used for the analysis of the differential tunneling conductance spectra shown exemplarily for top UC FeTe layer spectra.** (**A**) Raw data of d$I$/d$U$ spectra measured at temperatures of $T$ = 1.1 K, i.e. below the critical temperature $T_c$, and at $T$ = 8 K, i.e. above $T_c$, as indicated. Measurement parameters: $U$ = 10 mV, $I_s$ = 200 pA, $U_{mod}$ = 0.15 mV. (**B**) Background correction: the spectrum taken at $T$ = 1.1 K was divided by the spectrum measured at $T$ = 8 K in order to correct the non-constant background sample density of states in the normal state. (**C**) Slope correction: a constant slope has been subtracted in order to have d$I$/d$U$(-10mV) = d$I$/d$U$(+10mV). (**D**) Symmetrization with respect to zero bias: average of the original spectrum from (C) and its mirror with respect to zero bias. (**E**) Normalization: division by d$I$/d$U$(±10mV). (**F**) Fit of the measured spectrum to BCS theory using the Dynes function $N(E) = N_n(E_F) \cdot \Re \left[ \frac{|E|+i\Gamma(T)}{\sqrt{(|E|+i\Gamma(T))^2 - \Delta^2(T)}} \right]$ for the density of states $N(E)$ of the sample as described in [1]. Here, $N_n(E_F)$ is the sample density of states in the normal state which is assumed to be energy independent after normalization, $\Re$ denotes the real part, $\Gamma(T)$ is the resulting temperature dependent quasiparticle lifetime broadening factor, and $\Delta(T)$ is the resulting temperature dependent energy gap.

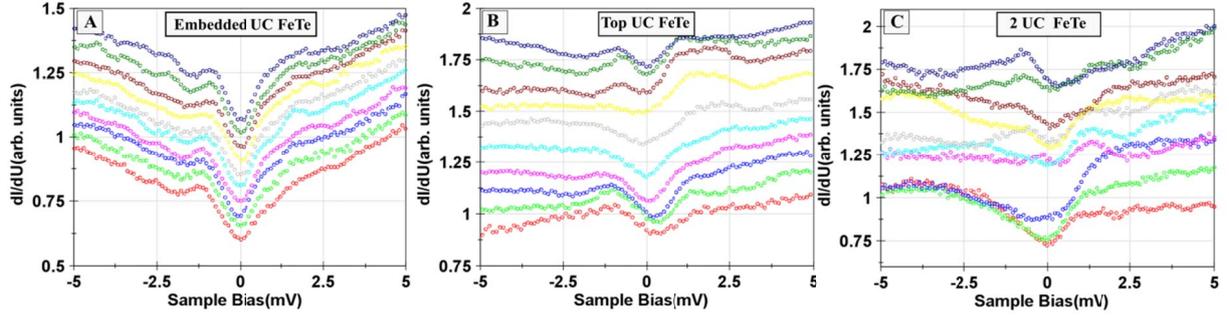

**Fig. S4: Spatial variation of differential tunneling conductance spectra measured at T = 1.1 K on FeTe layers of different thickness grown on $Bi_2Te_3$.** **(A)** Spectra on the embedded UC FeTe have been taken on 10 points along a line of 1 nm length ($U$ = 6 mV, $I_s$ = 300 pA, $U_{mod}$ = 0.2 mV). **(B)** Spectra on the top UC FeTe have been taken on 10 points along a line of 2 nm length ($U$ = 10 mV, $I_s$ = 600 pA, $U_{mod}$ = 0.25 mV). **(C)** Spectra on the 2 UC thick FeTe have been taken on 10 points along a line of 4 nm length ($U$ = 5 mV, $I_s$ = 300 pA, $U_{mod}$ = 0.2 mV). All spectra have been normalized to the average of the d$I$/d$U$ values at +5 mV and -5 mV, and are vertically shifted for clarity.

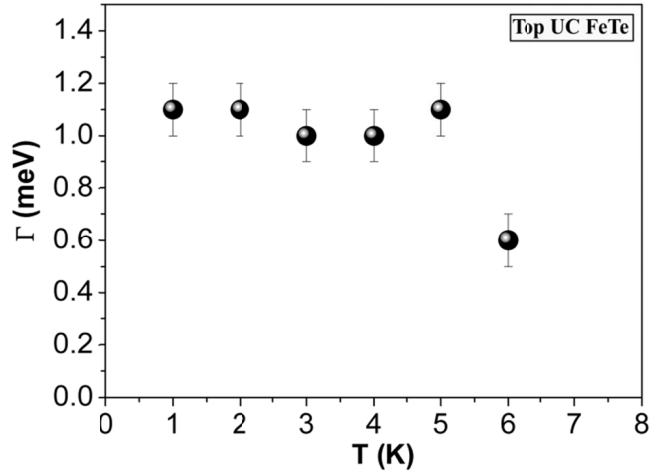

**Fig. S5: Temperature dependence of the lifetime broadening factor resulting from the fitting of the spectra in Fig. 2(E) of the main manuscript.**

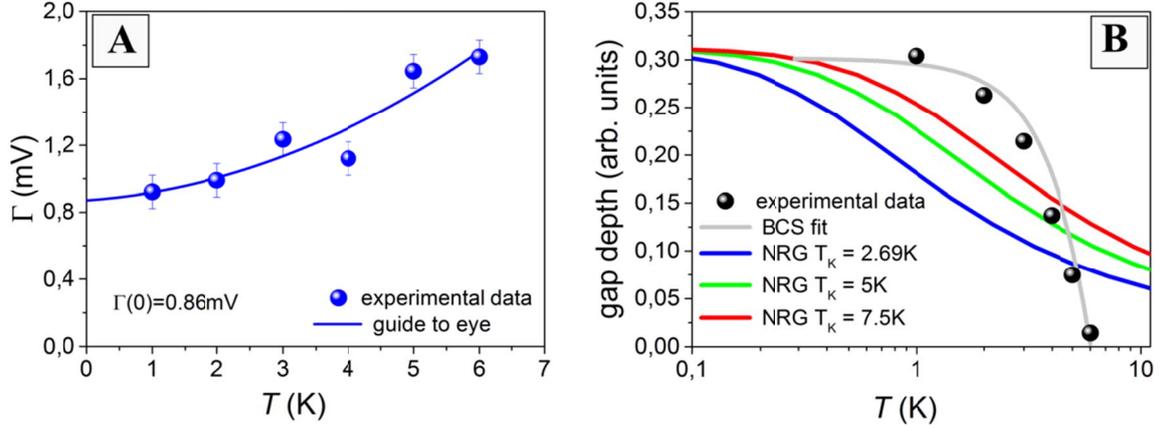

**Fig. S6: Comparison of the analysis of the data in Fig. 2(E) of the main text within a Kondo model, and the analysis within BCS theory.** In order to rule out, that the spectroscopic gap-feature at zero bias, which was observed for the thin FeTe films on $Bi_2Te_3$, is due to a Kondo lattice behavior of the Fe atoms [2], the data of Fig. 2(E) has been analyzed within a Kondo model. **(A)** The $dI/dU$ spectra in Fig. 2(E) (before background correction and symmetrization) have been fitted to Fano-functions $\frac{dI}{dU}(U) \propto c_1 + c_2 \frac{(q+\varepsilon)^2}{1+\varepsilon^2}$, with $\varepsilon = \frac{U-U_K}{\Gamma(T)}$ [3]. Here, $c_1$ and $c_2$ are offsets and amplitudes of the spectra, respectively, $q$ is the so-called form factor, $U_K$ is the voltage position, and $\Gamma(T)$ is the half-width at half maximum of the Kondo resonance. The resulting temperature dependence of $\Gamma$, which is shown in (A), interpolates to a value of $\Gamma(0) \approx 0.86$ mV, which would correspond to a Kondo temperature of $T_K = \frac{0.27\Gamma(0)}{k_B} \approx 2.69$ K, using Wilson's definition of the Kondo temperature [4]. **(B)** The temperature evolution of the extracted gap depth of the experimental data (markers) is compared to NRG calculations for a spin-1/2 Kondo impurity in the strong coupling regime [5] assuming different Kondo temperatures (colored lines). Obviously, the temperature dependence of the gap depth does not fit to that of a Kondo lattice, independent of the choice of $T_K$. In contrast, it fits well to the temperature dependence of the gap depth of a superconductor extracted from the BCS model (grey curve). The depth values at different temperatures for the grey curve were obtained employing a Dynes density of states with $\Delta$ as calculated from BCS theory [1,6].

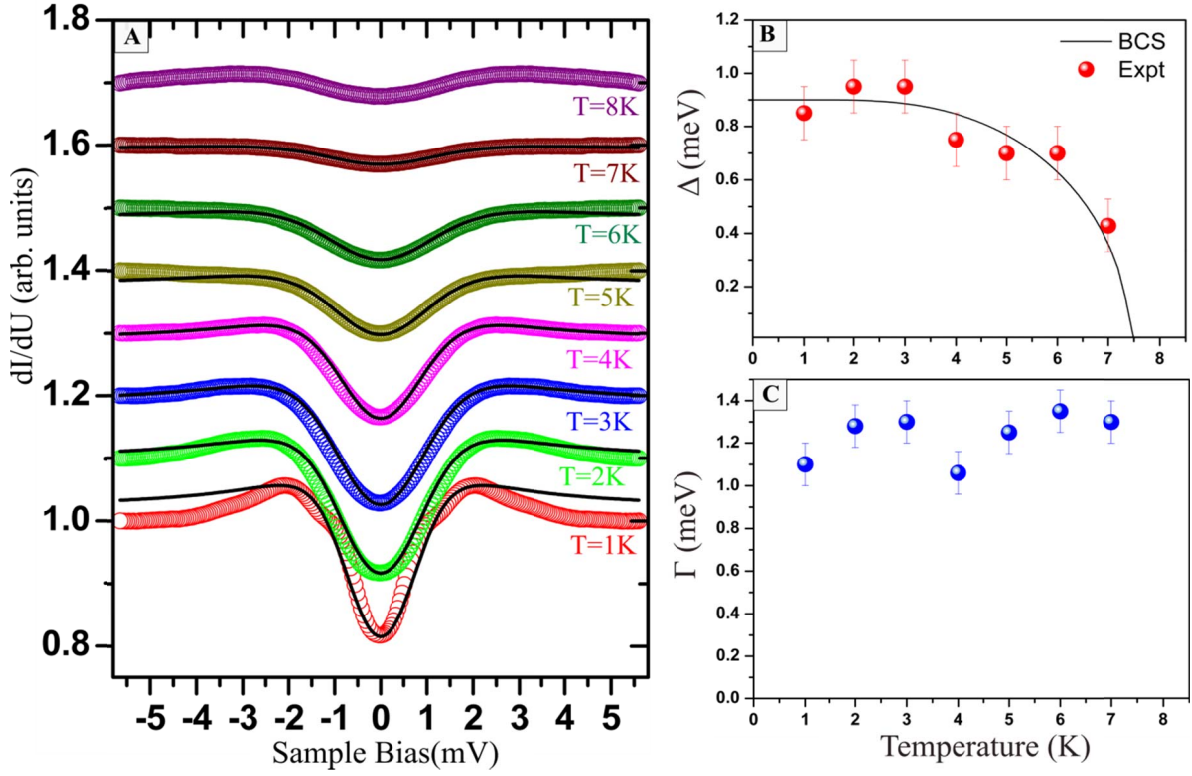

**Fig. S7: Temperature dependence of the differential tunneling conductance spectra measured on the embedded UC. (A)** Background corrected (using the data measured at $T = 10$ K) and symmetrized d$I$/d$U$ spectra (colored markers) measured at different temperatures as indicated. The spectra are spatially averaged over an area of 1 nm$^2$. Measurement parameters: $U = 6$ mV, $I_s = 300$ pA, $U_{\text{mod}} = 0.2$ mV. The lines show fits to BCS theory using the Dynes function (same as in caption of Fig. S3). **(B)** Temperature dependence of the gap values resulting from the fits in (A) (markers). The line shows the fit to the temperature dependence expected from BCS theory (see caption of Fig. S6) resulting in $\Delta(0) = 0.9$ meV and $T_c = 7.5$ K. **(C)** Lifetime broadening factors resulting from the fits in (A).

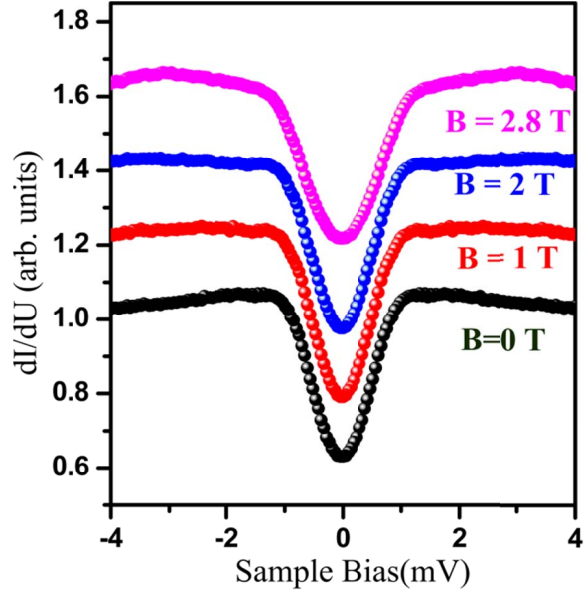

**Fig. S8: Magnetic field dependence of differential tunneling conductance spectra measured at $T = 1.1$ K on the embedded UC.** The magnetic field $B$ as indicated is applied perpendicular to the sample surface. The data is background corrected and symmetrized. Measurement parameters: $U = 6$ mV, $I_s = 300$ pA, $U_{mod} = 0.2$ mV. There is basically no change up to $B = 2$ T, and only a slight broadening of the gap at $B = 2.8$ T.

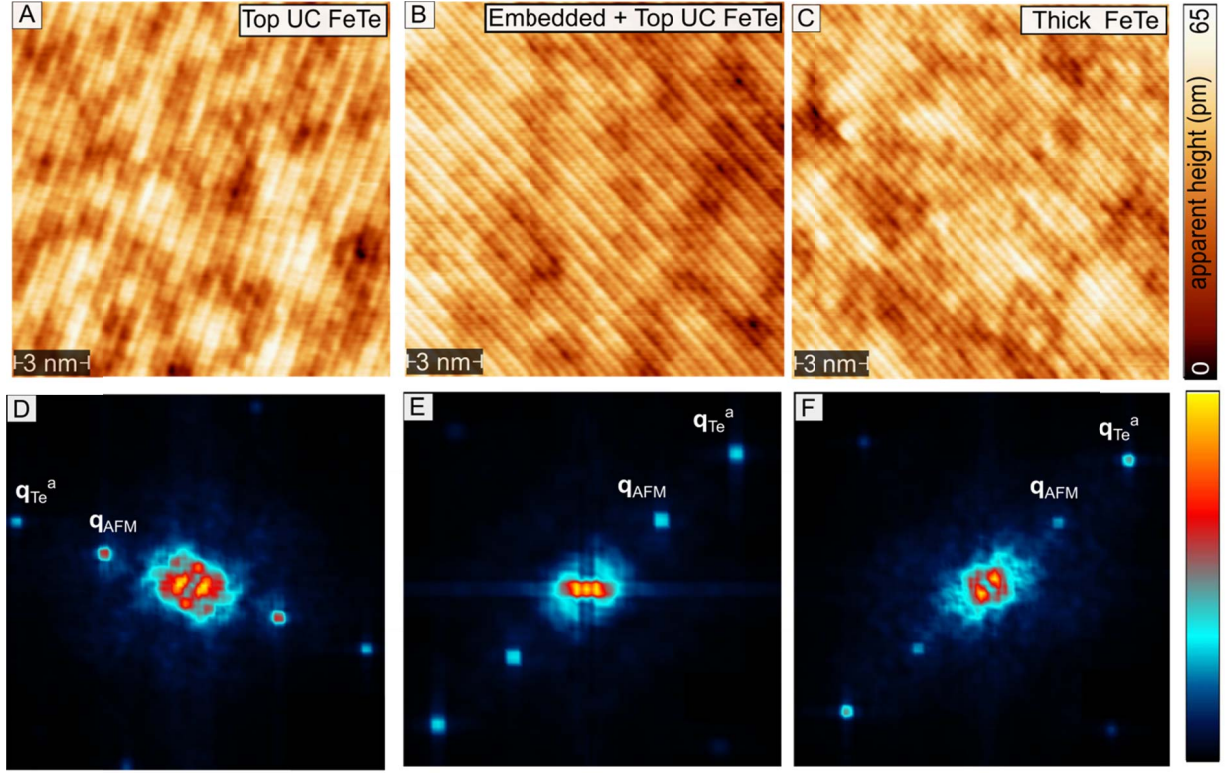

**Fig. S9: Spin-resolved STM topographs (A-C) taken at $T$ = 1.1 K of FeTe films of different thicknesses, as indicated, grown on $Bi_2Te_3$, and corresponding FFT images (D-F).** The spin contrast is visible in the topographs by a superstructure with twice the atomic lattice periodicity, and in the corresponding FFT images by the spots at wave vector $q_{AFM}$ = 1/2 $q_{Te}^a$. The intensity of the spin contrast is comparable for all three thicknesses. Measurement parameters of the topographs: (**A**) $U$ = 100 mV, $I_s$ = 100 pA, $B$ = 0.5 T; (**B**) $U$ = 200 mV, $I_s$ = 100 pA, $B$ = 1 T; (**C**) $U$ = 70 mV, $I_s$ = 100 pA, $B$ = 1 T. Image size of all FFTs is 0.54 Å$^{-1}$.

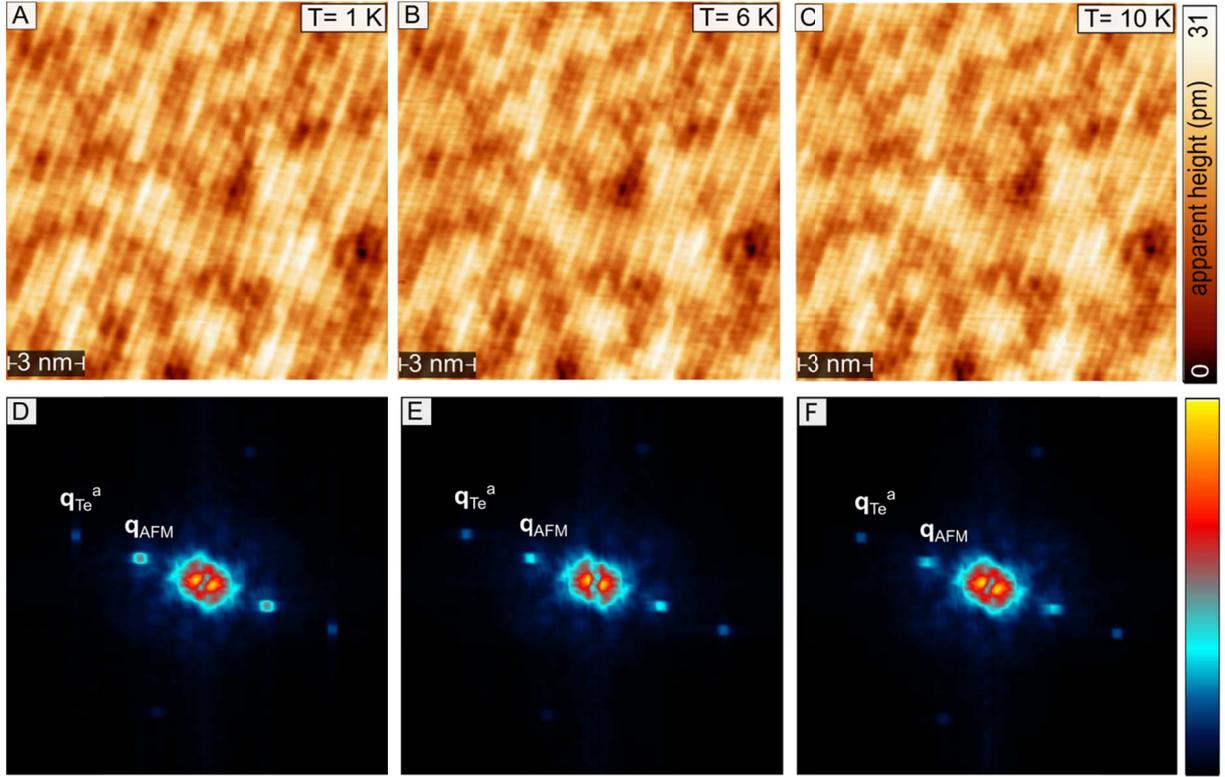

**Fig. S10: Spin-resolved STM topographs (A-C) of top UC FeTe film taken at different temperatures, as indicated, and corresponding FFT images (D-F).** The position of the surface area is the same in all three topographs. The spin contrast is clearly visible in the topographs by the superstructure with twice the atomic lattice periodicity, and in the corresponding FFT images by the spots at wave vector $\mathbf{q}_{AFM}= 1/2\ \mathbf{q}_{Te}^{a}$. The intensity of the spin contrast is almost the same at all three temperatures. Similar images at other temperatures have been taken in order to quantify the temperature dependence of the spin contrast via the ratio of intensities of the FFT spots at $\mathbf{q}_{AFM}$ and $\mathbf{q}_{Te}^{a}$, which is plotted in Fig.4 (F) of the main manuscript. Measurement parameters of the topographs: $U = 100$ mV, $I_s = 100$ pA, $B = 0.5$ T. Image size of all FFTs is 0.75 Å$^{-1}$.

**Supplementary References:**